\documentclass[twocolumn]{revtex4}

\usepackage{graphicx}
\usepackage{dcolumn}
\usepackage{bm}
\begin{document}

\title{Effect of resistance feedback on spin torque-induced switching of nanomagnets}

\author{Samir Garzon$^1$}
\author{Richard A. Webb$^1$}
\author{Mark Covington$^2$}
\author{Shehzaad Kaka$^2$}
\author{Thomas M. Crawford$^1$}

\affiliation{
$^1$Department of Physics and Astronomy and USC Nanocenter, University of South Carolina, Columbia, SC 29208, USA.\\
$^2$Seagate Research, 1251 Waterfront Place, Pittsburgh, PA 15222,
 USA.\\}

\date{\today}

\begin{abstract}
In large magnetoresistance devices spin torque-induced changes in resistance can produce GHz current and voltage oscillations which can affect magnetization reversal. In addition, capacitive shunting in large resistance devices can further reduce the current, adversely affecting spin torque switching. Here, we simultaneously solve the Landau-Lifshitz-Gilbert equation with spin torque and the transmission line telegrapher's equations to study the effects of resistance feedback and capacitance on magnetization reversal of both spin valves and magnetic tunnel junctions. While for spin valves parallel (P) to anti-parallel (AP) switching is adversely affected by the resistance feedback due to saturation of the spin torque, in low resistance magnetic tunnel junctions P-AP switching is enhanced. We study the effect of resistance feedback on the switching time of MTJ's, and show that magnetization switching is only affected by capacitive shunting in the pF range.
\end{abstract}

\maketitle

\section{Introduction}
The use of spin-transfer torque~\cite{slonczewski_JMMM1996,berger_PRB1996} for controlling the resistance state of nanoscale magnetic tunnel junctions (MTJ's) has the potential for becoming the paradigm of non-volatile memory technology~\cite{chappert_NatMat2007}. However, the main challenges that must be overcome to implement spin transfer torque random access memory (STT-RAM) are (i) increasing the voltage swing to achieve integration into traditional GHz sensing circuitry, and (ii) reducing the critical current required to switch MTJ's to a level which makes them compatible with current CMOS technology~\cite{ikeda_IEEETED2007}. MgO MTJ's fabricated with CoFeB low magnetization materials have shown tunneling magnetoresistance (TMR) exceeding 100$\%$ for devices with resistance in the k$\Omega$ range and switching currents $\sim$MA/cm$^2$~\cite{diao_JPhysCondMat2007,diao_APL2005,ikeda_JMMM2007}. These devices not only exceed the required voltage swing threshold, but also reduce the critical switching current density $J_C$ to within the order of magnitude needed for CMOS integration. Current efforts are aimed at reducing the resistance-area (RA) product of high TMR MTJ's in order to decrease power dissipation and noise~\cite{isogami_APL2008}. In addition, to further reduce $J_C$ at reasonably fast pulsewidths without sacrificing thermal stability~\cite{urazhdin_PRL2003,li_PRB2004}, material development and careful device design are necessary. However, typical models of spin-torque-induced magnetization reversal ignore the changes in device resistance that occur during the switching process~\cite{sun_PRB2000}. For large TMR devices the large change in resistance between antiparallel (AP) and parallel (P) states ($\Delta R=R_{AP}-R_P$) invalidates the constant resistance approximation. Furthermore, at these large resistances capacitive shunting decreases the effective current through the MTJ, affecting the switching process. Therefore, understanding the effect of capacitance and time-dependent resistance on magnetization dynamics and switching is relevant for future device design.

As an alternative to MTJ's, metallic spin valves with very low resistance ($\lesssim$1 $\Omega$), could also be used in memory storage applications. Even though the voltage swing is rather limited in these type of devices due to the low values of magnetoresistance (MR) typically obtained, spin valves are still attractive due to their lower power consumption. Furthermore, as the dimensions of the non-essential non-magnetic layers of spin valves are decreased, the MR could increase by an order of magnitude, thus invalidating the constant resistance approximation. It is therefore useful to understand the effects of time-dependent resistance oscillations on both metallic spin valve and MTJ switching.

\section{Model}

\begin{figure*}[hbt] 
\begin{center} 
\includegraphics[width=6.5in]{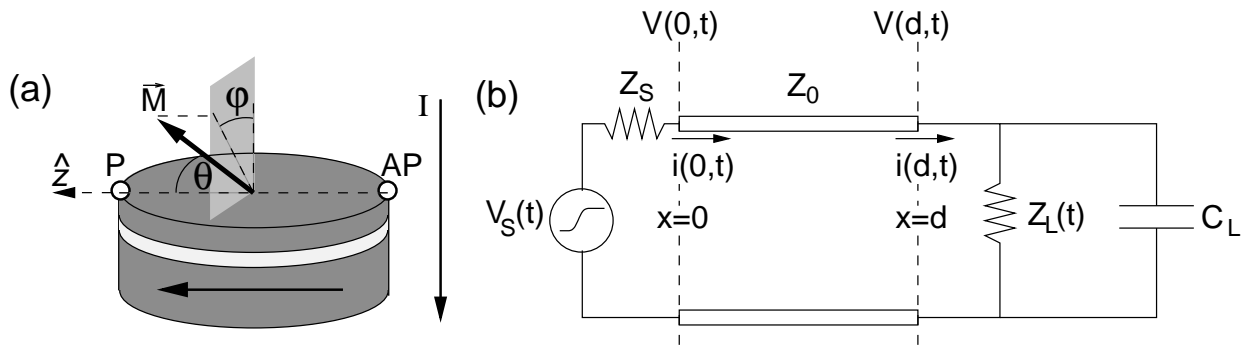} 
\caption{(a) Schematic of nanopillar MTJ or spin valve. Two ferromagnetic layers (dark gray) are separated by a thin spacer layer (light gray), which can be either insulating (for MTJ's) or metallic (for spin valves). The bottom ``reference layer'' has a fixed magnetization, while the magnetization $\vec{M}$ of the top ``free layer'' can be described by the polar angle $\theta$ with the $\hat{z}$ axis and the azimuthal angle $\phi$. The P and AP stable points of $\vec{M}$, which lie along the easy axis (dashed line) at $\theta=0$ and $\theta=\pi$ respectively, are represented by open circles. The direction of positive current is shown. (b) Electrical setup. A device with resistance $Z_L(t)$ and shunting capacitance $C_L$ is connected via a transmission line of characteristic impedance $Z_0$ and length $d$ to a voltage source $V_S(t)$ with internal impedance $Z_S$.}\label{fig:figLayout} \end{center} 
\end{figure*}

We consider a typical MTJ or metallic spin valve patterned into an elliptical nanopillar~\cite{katine_PRL2000} as shown in Fig.~\ref{fig:figLayout}(a). The bottom ``reference layer'' has a fixed magnetization and is separated from the top ``free layer'' by a thin spacer, which is insulating for MTJ's and metallic for spin valves. The geometry of the free layer induces strong demagnetizing fields which are the main source of magnetic anisotropy, and thus the magnetic energy is minimized whenever the free layer magnetization $\vec{M}$ points along the easy axis direction (towards P or AP). The spin torque generated by a current (voltage) pulse induces precession of $\vec{M}$, thereby changing the resistance of the trilayer due to giant magnetoresistance (GMR). As the device resistance changes, the actual current and voltage at the device also change due to a time dependent impedance mismatch with the transmission line. Spin torque switching is usually modeled assuming that during the magnetization reversal process the device resistance is constant, ignoring the effect of resistance feedback. Furthermore, the device is typically considered to be purely resistive, ignoring the effects of capacitive shunting. Here we simultaneously solve the Landau-Lifshitz-Gilbert (LLG) equation including Slonckzewski's spin transfer torque in the macrospin approximation~\cite{sun_PRB2000}, together with the telegrapher equations~\cite{pozar_1997} for the lossless transmission line connected to the device~\cite{branin_IEEE1967}. We use different forms for the angular dependence of the spin torque efficiency $g(\theta,p)$, where $p$ is the ferromagnet polarization. As shown in Fig.~\ref{fig:figLayout}(b) the electrical setup consists of a voltage source with impedance $Z_S$ which produces a waveform $V_S(t)$, a transmission line of length $d$ with characteristic impedance $Z_0=\sqrt{L/C}$ ($L$ and $C$ are the inductance and capacitance per unit length), and a spin valve/MTJ device which can be considered as a lumped element with resistance $Z_L(t)$ in parallel with a constant capacitance $C_L$. By following Branin's method of forward and reflected waves that relates the voltages and currents at the ends of a transmission line~\cite{branin_IEEE1967}, the boundary conditions $V(0,t)=V_S(t)-i(0,t) Z_S$ (at the source output) and $V(d,t)=\left[ i(d,t)-C_L \partial V(d,t)/\partial t \right]  Z_L(t)$ (at the device), lead to the equations

\begin{eqnarray}
\label{equ:voltage1}
V(0,t)&=&\frac{1}{1+Z_0/Z_S}\left\{V(d,t-\tau)\left[1-\frac{Z_0}{Z_L(t-\tau)}\right]\right.\nonumber\\
&-&\left.\frac{\partial V(d,t-\tau)}{\partial t} Z_0 C_L+V_S(t)\frac{Z_0}{Z_S}\right\}
\end{eqnarray}



\noindent and

\begin{eqnarray}
\label{equ:voltage2}
V(d,t)&=&\frac{1}{1+Z_0/Z_L(t)}\left\{V(0,t-\tau)\left[1-\frac{Z_0}{Z_S}\right]\right.\nonumber\\
&-&\left.\frac{\partial V(d,t)}{\partial t} Z_0 C_L+V_S(t-\tau)\frac{Z_0}{Z_S}\right\}
\end{eqnarray}


\noindent for the voltages at the source and device respectively, where $\tau=\sqrt{LC}d$ is the time for an electrical signal to propagate from the source to the device along the transmission line. Using the initial conditions $V(0,t<0)=0$ and $V(d,t<\tau)=0$, the voltage and current at the device can be found at any time $t>0$ for a given $V_S(t)$ if $Z_L(t)$ is known. However, the resistance of the device,

\begin{equation}
\label{equ:impedance}
Z_L(t)=R_P+\left(R_{AP}-R_P\right)\sin^2\frac{\theta(t)}{2},
\end{equation}

\noindent depends on the angle $\theta$ between the magnetization of the free and pinned layers [Fig.~\ref{fig:figLayout}(a)], which is driven by the spin torque. The MR and TMR are defined, as usual, as $(R_{AP}-R_P)/R_P$. Given $\theta(t)$ and $\phi(t)$, which completely describe the nanomagnet state at time $t$ in the macrospin model [Fig.~\ref{fig:figLayout}(a)], we use Eqs. (\ref{equ:voltage1})-(\ref{equ:impedance}) and the boundary conditions to find $i(d,t)$. We then numerically calculate $\theta(t+dt)$ and $\phi(t+dt)$ using a standard Runge-Kutta method~\cite{press_2002}. We consider thermal effects by finding a set of initial orientations $\theta_0$, $\phi_0$ representative of a given temperature $T$. For single trajectory simulations we assume completely deterministic evolution of the free layer magnetization, whereas for calculating average switching times we include the effect of random thermal fluctuations by solving the stochastic LLG equation~\cite{garzon_PRB2009}.

\section{Results and discussion}

\begin{figure*}[t]
\begin{center}
\includegraphics[width=6.5in]{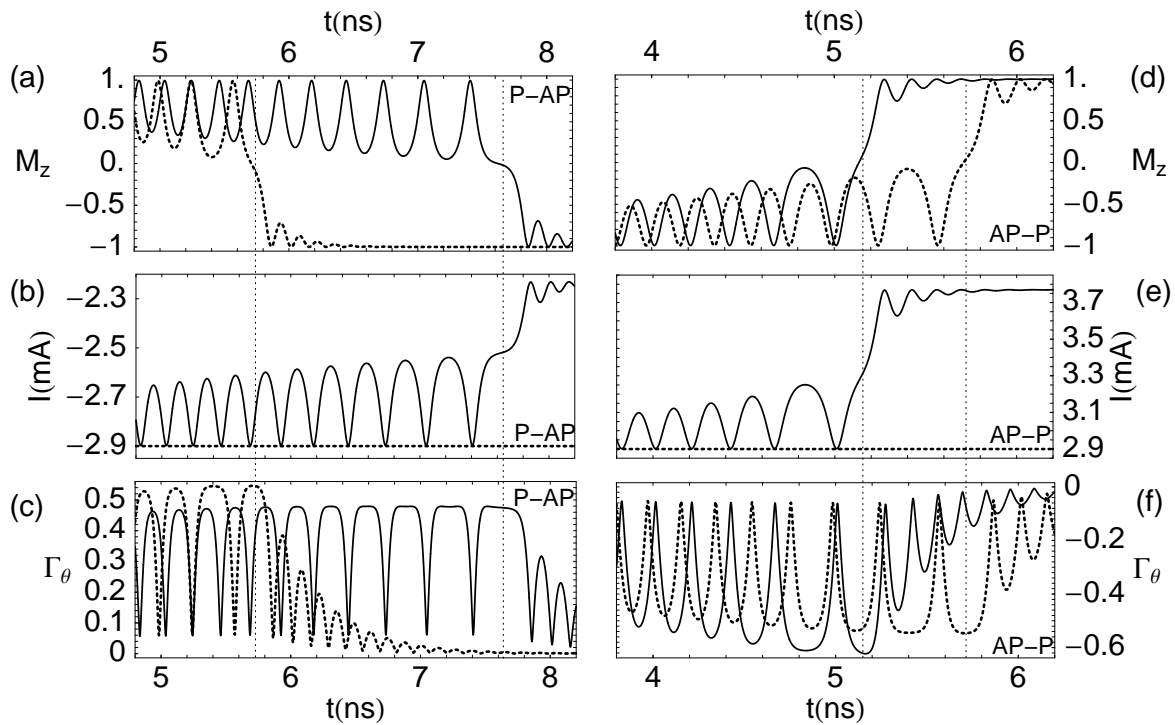}
\caption{(a) Easy axis component of the magnetization, M$_Z$, (b) current, and (c) spin torque as a funtion of time for P-AP switching, including (solid line) and excluding (dashed line) RIF for $g(\theta,p)=$0.48. (d)-(f) Similar plots as (a)-(c) but for AP-P switching. Device parameters are $R_P$=0.2$\Omega$ and MR=60$\%$.
}\label{fig:figTraj}
\end{center}
\end{figure*}

\subsection{Resistance feedback}

We will first discuss the case $C_L=0$. For simplicity we only consider $Z_S=Z_0$ such that the current and voltage at the device are given by

\begin{equation}
\label{equ:i}
i(d,t)=V_S(t-\tau) \frac{1}{Z_0+Z_L(t)},
\end{equation}

\noindent and

\begin{equation}
\label{equ:V}
V(d,t)=V_S(t-\tau) \frac{Z_L(t)}{Z_0+Z_L(t)}.
\end{equation}

\noindent Since in this case the source is perfectly matched to the transmission line, all the power reflected at the device will be absorbed by the source without any further reflections, and thus the only effect of the transmission line length is to introduce a delay in the device response. Equations \ref{equ:i} and \ref{equ:V} show that changes in the device impedance will affect both the device current and voltage. However, device current and voltage are affected rather differently by changes in device resistance. In the limit where $Z_L(t)\gg Z_0$ the current is inversely proportional to $Z_L(t)$, whereas the voltage is independent of $Z_L(t)$. In the opposite limit the current becomes independent of $Z_L(t)$ while the voltage is directly proportional to $Z_L(t)$. The different behavior of current and voltage is of great importance for understanding the difference in the resistance feedback effect for metallic spin valves and magnetic tunnel junctions. 

For metallic spin valves the magnitude of the spin torque is given by $\tau_s=\hbar/2e\phantom{.} g(\theta,p) \phantom{.} i(d,t) \phantom{.} sin \theta$ where the function $g(\theta,p)$ describes the angular dependent efficiency. Since the spin torque is proportional to the device current, Eq.~\ref{equ:i} shows that for devices with $Z_L(t)\ll Z_0$ the spin torque is independent of $Z_L(t)$ and thus in this limit changes in device resistance do not affect spin torque switching. This is the typical situation for metallic spin valves driven by 50 $\Omega$ sources. However, if spin valves are driven by low impedance sources such that $Z_L(t)\sim Z_0$ then Eq.~\ref{equ:i} shows that the current, and therefore the spin torque, will depend inversely on the device resistance. We call this effect resistance-current feedback (RIF). 

In contrast with metallic spin valves, theoretical models~\cite{slonczewski_JMMM2007,sun_JMMM2008,heiliger_PRL2008} and experimental evidence~\cite{kubota_NATPHYS2008,sankey_NATPHYS2008} suggest that in magnetic tunnel junctions the spin torque is a function of voltage and does not depend on $Z_L(t)$, the time dependent device resistance, but only on $R_P$. Therefore we assume that the spin torque in MTJ's is given by

\begin{equation}
\label{equ:torque}
\tau_s=\frac{\hbar}{2e} \frac{p}{1+p^2} \frac{V(d,t)}{R_P} \sin\theta,
\end{equation}

\noindent where $V(d,t)$ is the device voltage~\cite{slonczewski_JMMM2007,sun_JMMM2008}. Although in general the torque can be a more complicated function of voltage, experiments have shown that at small bias the parallel torque is linear in voltage~\cite{kubota_NATPHYS2008,sankey_NATPHYS2008}. For large resistance MTJ's such that $Z_L(t)\gg Z_0$, Eq.~\ref{equ:V} shows that the voltage is independent of $Z_L(t)$ and thus the spin torque is unchanged by any changes in the device resistance. However, for MTJ's with RA products $\sim$1$\Omega\mu m^2$, which are of great interest for applications, $Z_L(t)$ is comparable to $Z_0$. For this type of devices a decrease in $Z_L(t)$ leads to smaller device voltage and therefore smaller spin torque. We call this effect resistance-voltage feedback (RVF). In the next two sections we will study the resistance feedback in spin valves (RIF) and in MTJ's (RVF).

\subsubsection{Spin valves}

The situation envisioned here consists of a large MR (60$\%$), low resistance ($R_P$=0.2 $\Omega$) spin valve which is directly connected to a local switchable voltage source (e.g. a transistor) with very low output impedance $\sim R_P$, as in typical magnetic random access memory (MRAM) architectures ~\cite{ikeda_IEEETED2007}. We compare simulations of spin-torque driven magnetization switching which ignore and include resistance-current feedback. As a source we use a voltage step

\begin{equation}
\label{equ:step} V_S(t)=\left\{
\begin{array}{c l}
A[1-\exp(-t^2/2\sigma^2)] & t>0\\
0 & t\leq0
\end{array}\right.,
\end{equation}

\noindent of amplitude $A$ and risetime $t_{10-90\%}\approx1.7\sigma$. For all the results shown in this section we use the electrical parameters $Z_S=Z_0=$0.2$\Omega$ and $t_{10-90\%}\approx$34 ps, and use the device parameters from our Co/Cu/Co devices~\cite{garzon_PRB2008} which have an ellipsoidal free layer with dimensions 125$\times$75$\times$2 nm$^3$ and magnetization $M_S$=1.01$\times$10$^6$A/m. The uniaxial and uniplanar anisotropy fields used for the simulations result from demagnetizing effects due to shape anisotropy. For a Co/Cu/Co spin valve structure, assuming $p$=0.48 for Co~\cite{bass_JMMM1999}, the Valet-Fert description gives a maximum MR of 60$\%$. We first assume a spin torque term of the Slonczewski form but with constant $g(\theta)$=0.446, which is the average value of Slonczewski's ballistic $g(\theta,p=0.48)$ in the range $0<\theta<\pi$~\cite{slonczewski_JMMM1996}.

Fig.~\ref{fig:figTraj}(a) shows the component of the free layer magnetization along the easy axis direction, $M_Z$, as a function of time including (solid line) and ignoring (dashed line) resistance feedback for an initial orientation of $\vec{M}$ with $\theta=0.3$, $\phi=\pi/2$ for a pulse with A=-1.16 mV (negative polarity is required for P-AP switching). For short times (not shown), when $\theta\ll1$, the two traces overlap, since to first order in $\theta$ the device resistance is equal to $R_P$. However, as the amplitude of precession increases, the device resistance starts to deviate from $R_P$ resulting in slower build up of oscillations when RIF is considered, and leading to delayed switching of the MTJ.
For the case illustrated here, more than three full additional precession cycles (i.e. more than six $M_Z$ oscillations) are required before switching occurs. As shown in Figs.~\ref{fig:figTraj}(b) and (c), the slower build-up of the oscillations in M$_Z$ occurs since RIF reduces the current through the device [Fig.~\ref{fig:figTraj}(b)], limiting the magnitude of the spin torque [Fig.~\ref{fig:figTraj}(c)]. Note that when RIF is ignored, the amplitude of the spin torque oscillations increases up to the point where switching occurs. On the other hand, when RIF is considered, the amplitude of the spin torque oscillations saturates many cycles before switching. Since we assume $g(\theta)=$const, Slonczewski's spin torque term predicts increasing spin torque amplitude for increasing $\theta$ up to the switching point at $\pi/2$. However, the current decreases for increasing $\theta$ due to increased device resistance. Therefore the spin torque is proportional to $\sin\theta/(1+\textbf{TMR}\sin^2\theta/2)$, and has a maximum at $\cos\theta=\textbf{TMR}/(2+\textbf{TMR})$, explaining the observed saturation. 

\begin{figure}[b]
\begin{center}
\includegraphics[width=3.25in]{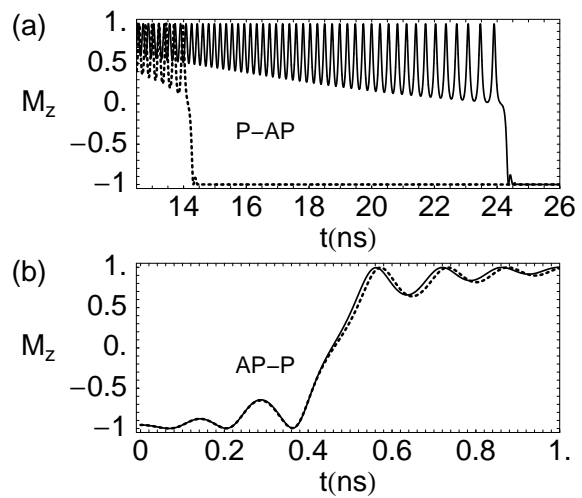}
\caption{Easy axis component of the magnetization, M$_Z$, as a function of time for (a) AP-P and (b) P-AP switching, including (solid line) and excluding (dashed line) RIF. Here we use the ballistic form of $g(\theta,p)$ proposed by Slonczewski and the same device parameters as in Fig. 2.}\label{fig:figslonc}
\end{center}
\end{figure}

Since AP-P switching does not occur for A=1.16 mV, we increase the pulse amplitude to A=1.508 mV in order to obtain the same initial device current of 2.9 mA. For AP-P switching, Fig.~\ref{fig:figTraj}(d) shows that RIF effects produce earlier magnetization reversal. However, the decrease in the AP-P switching time due to RIF is noticeably smaller than the increase in P-AP switching time. The reasons for this behavior are twofold: first, the relative modulation in the current is smaller for AP-P switching [Fig.~\ref{fig:figTraj}(e)] since the fractional change in resistance $\Delta R/R_{AP}$ (relevant for AP-P) is smaller than $\Delta R/R_{P}$ (relevant for P-AP). And second, the amplitude of the spin torque does not saturate but increases monotonically up to the switching point [Fig.~\ref{fig:figTraj}(f)], since in this case the magnitude of the current increases as $\theta$ approaches $\pi/2$, and thus there are no spin-torque critical points in the $\pi/2<\theta<\pi$ range. By choosing the same large voltage (A=1.508 mV) for both P-AP and AP-P switching (data not shown), we observe that RIF decreases the difference between P-AP and AP-P switching times, making the switching process more symmetric. However, for a given voltage step magnitude, P-AP switching always occurs at shorter times. 

We have also studied the effect of including a more realistic $g(\theta,p)$ such as that proposed by Slonczewski~\cite{slonczewski_JMMM1996}. First, we used the values P=0.48 and A=1.508 mV, as above, but only AP-P switching occured (results not shown). We then increased the pulse amplitude (A=2.45 mV) to make the average spin 
torque amplitude in the range $0<\theta<\pi/2$ match the amplitude of the spin torque used before for $g(\theta)$=0.446. $M_Z$ as a function of time is shown in Fig.~\ref{fig:figslonc} for both (a) P-AP and (b) AP-P switching with (solid line) and without (dashed line) RIF. Since in this case the average amplitude of the spin torque in the range $0<\theta<\pi/2$ is more than 3 times smaller than in the range $\pi/2<\theta<\pi$, AP-P switching occurs extremely fast, within 0.5 ns, while AP-P switching without RIF occurs only after 14 ns. Therefore the effects of RIF for AP-P switching are almost negligible, whereas for P-AP the switching time increases by more than 70$\%$. By comparing the average value of $g(\theta,p)$ during P-AP and AP-P switching, and using the maximum value of MR for a given spin polarization (using the Valet-Fert model), it can be argued that Slonczewski's ballistic $g(\theta,p)$ together with RIF will always lead to a situation where AP-P switching occurs much earlier and requires lower pulse amplitudes than P-AP switching.

\subsubsection{Magnetic tunnel junctions}

\begin{figure*}[t]
\begin{center}
\includegraphics[width=6in]{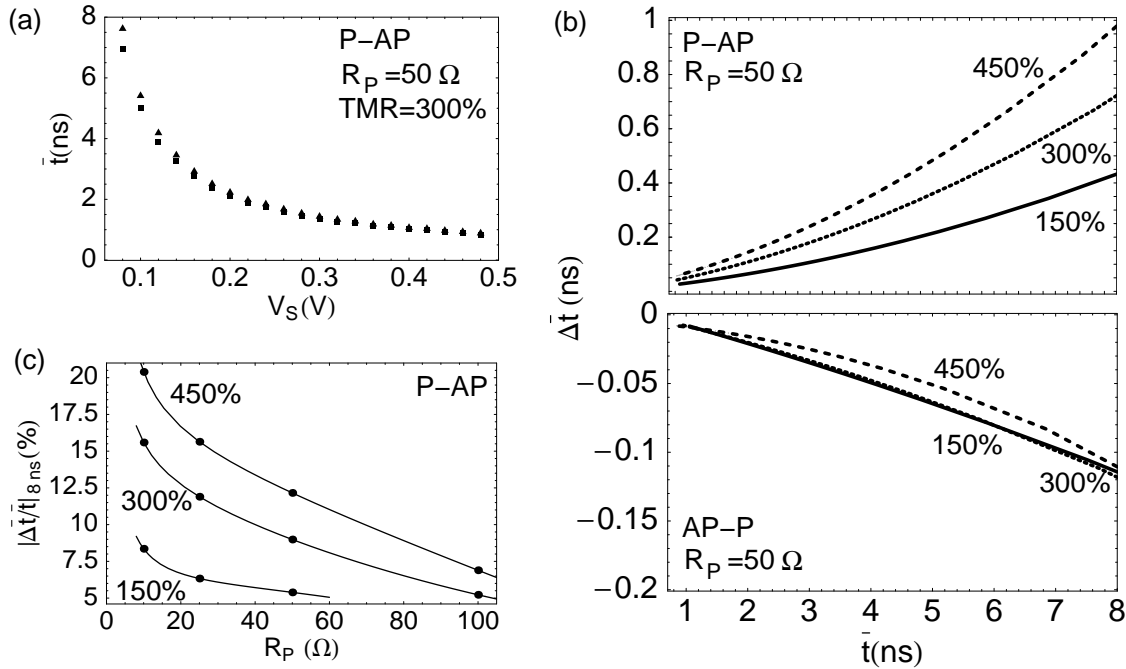}
\caption{(a) Average P-AP switching time $\bar t$ as a function of pulse voltage for a device with $R_P$=50$\Omega$ and TMR=300$\%$ with (squares) and without (triangles) RVF. (b) Difference in average switching time due to RVF, $\Delta \bar t= \bar t-\bar t_{RVF}$, as a function of $\bar t$ for a device with $R_P$=50$\Omega$ and TMR=150, 300, and 450 $\%$. The upper (lower) panel shows results for P-AP (AP-P) switching. In the lower panel the curves for different values of TMR overlap. (c) Relative change in P-AP switching time due to RVF at $\bar t$=8 ns as a function of $R_P$ for TMR=150, 300, and 450 $\%$.}\label{fig:tbar}
\end{center}
\end{figure*}

We now study the effect of resistance-voltage feedback on MTJ switching for a spin torque term given by Eq.~\ref{equ:torque}. We use device parameters corresponding to CoFeB MTJ's reported by Diao~\cite{diao_APL2005}, but assume smaller RA products of the order of 1 $\Omega\mu m^2$~\cite{isogami_APL2008}. Whereas trajectories such as those in Fig.~2 show that for spin valves RIF slows P-AP and enhances AP-P magnetization reversal, similar results for MTJ's show that resistance-voltage feedback produces the opposite effect. This is expected since during P-AP switching, the increase in $Z_L(t)$ leads to larger MTJ voltage and therefore larger spin torque (see discussion following Eq.~\ref{equ:torque}). On the other hand for spin valves, as the resistance increases, the device current and thus the spin torque decrease, slowing the switching process. One similarity between resistance feedback in spin valves and MTJ's is the saturation of the spin torque. In MTJ's this saturation occurs for AP-P switching and can be explained using Eqs.~\ref{equ:V} and \ref{equ:torque} by noticing that the spin torque is proportional to $\sin\theta\phantom{i}Z_L(t)/[Z_0+Z_L(t)]$, which has a maximum only for $\theta>\pi/2$.

So far we have discussed the detailed effects of resistance feedback on a single trajectory. However, what is typically measured experimentally is an average over many different trajectories resulting from the slightly different initial states of the free layer magnetization that arise due to thermal effects. Pulsed switching experiments, for example, measure the switching probability $P_S$ which results from multiple repetitions of the switching process under the same experimental conditions. To better understand the measurable effects of RVF, we first generate a set of initial orientations of $\vec{M}$ around the P or AP energy minimum with a Maxwell-Boltzmann distribution corresonding to a temperature $T$=293 K. We then find the average switching time $\bar t$ with and without RVF as a function of source voltage in the presence of random thermal fluctuations~\cite{garzon_PRB2009}. Figure~\ref{fig:tbar}(a) shows $\bar t$ for P-AP switching with (squares) and without (triangles) RVF for a device with $R_P$=50 $\Omega$ and TMR=300$\%$. We use values of $p$ extracted from the TMR using Julliere's relation~\cite{julliere_PhysLettA1975}. Our results indicate that the average switching time and the critical switching voltage, $V_C$, decrease due to RVF. We observe that the average switching time can be fitted extremely well by $\bar t=t_0(1+V_A/(V_S-V_C))$, where $t_0$, $V_C$, and $V_A$ are fitting parameters. Therefore, to remove the voltage dependence, we subtract fits to the data shown in Fig.~\ref{fig:tbar}(a) without and with RVF and plot $\Delta \bar t=\bar t-\bar t_{RVF}$ as a function of $\bar t$, as shown by the dotted line in the top panel of Fig.~\ref{fig:tbar}(b). By following a similar procedure we find $\Delta \bar t$ for P-AP and AP-P switching for a device with $R_P$=50 $\Omega$ and TMR=150$\%$, 300$\%$, and 450$\%$. For AP-P switching [lower panel of Fig.~\ref{fig:tbar}(b)], traces for different values of the TMR are very close to each other and even overlap. We point out that in this case the small differences in $\Delta \bar t$ between the traces are not significant, since they are comparable to the error originating from the fitting procedure. What is significant is that the increase in AP-P switching times due to RVF is much smaller than the decrease in P-AP switching times. This behavior occurs since, as explained before, the effect of RVF decreases with increasing device resistance. The AP-P switching process starts with the device in the high resistance state and thus the voltage modulation due to impedance mismatch is smaller than for P-AP switching. This is similar to the case of spin valves where the effects of RIF are larger for P-AP switching. From Fig.~\ref{fig:tbar}(b) we observe that as the TMR increases the effects of RVF on P-AP switching become larger: for devices with TMR=450$\%$ and $\bar t=$ 8 ns, RVF can reduce the average P-AP switching time by 1 ns, which is more than 10$\%$ of $\bar t$. We performed a similar analysis for devices with $R_P$=10, 25, and 100 $\Omega$ and observed the same qualitative behavior as seen for the 50$\Omega$ device. The effect of varying the resistance is demonstrated in Fig.~\ref{fig:tbar}(c), which shows $|\Delta \bar t/\bar t|$ at $\bar t$=8 ns as a function of $R_P$ for different values of the TMR. RVF decreases the switching time to a larger extent as the device resistance decreases, in contrast with metallic spin valves where the effect of RIF decreases with resistance. For a device with $R_P$=10$\Omega$ and TMR=450$\%$, resistance-voltage feedback lowers the average switching time by up to 20$\%$. This means that in large TMR, low resistance MTJ's, asymmetric switching (P-AP vs AP-P) is expected due to RVF.

So far, in order to illustrate the effect of RVF on MTJ switching, we have only used the simplest approximation for the spin torque term, which ignores (i) the voltage dependence of TMR, (ii) any explicit dependence of the spin torque on the time dependent device resistance, and (iii) an additional ``out-of-plane'' or ``field-like'' spin torque~\cite{theodonis_PRL2006,sankey_NATPHYS2008,kubota_NATPHYS2008,li_PRL2008} which has a considerable effect in MTJ's but not in metallic spin valves. We briefly comment on each of these approximations.

(i) In our simulations the source and transmission line are well matched to avoid multiple pulse reflections. Since the rise time of the pulse is much shorter than the pulse duration required for MTJ switching, the initial transient voltage can be safely ignored. However, due to resistance-voltage feedback, the precessing free layer moment modulates the device voltage, thus also modulating the voltage dependent TMR. We estimate from Eq.~\ref{equ:V} that for a device with $R_P$=50$\Omega$ and TMR=150$\%$ a pulse with 500 mV amplitude at the source will produce device voltages of 250 mV at $\theta=0$ and 290 mV at $\theta=\pi/4$. According to Fig. 2 of Diao et al.~\cite{diao_APL2005} this implies a TMR decrease from $\sim$115$\%$ to $\sim$110$\%$. The fractional decrease in TMR is less than 5$\%$ whereas the fractional increase in spin torque is $\sim$16$\%$, estimated from the voltage increase between $\theta$=0 and $\theta=\pi/4$. Therefore the main effect of RVF is the modulation of the spin torque and not the modulation of TMR. However, we expect that the voltage dependence of the TMR will decrease the observed effects of RVF, and therefore a complete model should include the voltage dependence of TMR.

(ii) We assumed that the spin torque is proportional to the device voltage divided by $R_P$ (Eq.~\ref{equ:torque})~\cite{slonczewski_JMMM2007,sun_JMMM2008}. This approximation could break down at large voltages or for nonsymmetric tunnel junctions, since in that case the spin torque could depend explicitly on $Z_L(t)$, the time dependent device resistance. Depending on the functional form of the spin torque, the qualitative effects of resistance feedback on MTJ switching could be very different from those reported here.

(iii) Following Li et al.~\cite{li_PRL2008} we have included a field-like spin torque which is proportional to $V|V|\sin \theta$. We assume that for a voltage bias of 1 V both the usual spin torque and the field-like spin torque have equal magnitudes. We observe the same general behavior described in Fig. 4. This occurs since the field-like torque only becomes important at large voltages, whereas RVF has its largest effect at lower voltages (i.e. longer switching times).

\subsection{Effect of capacitive shunting}

\begin{figure}[t]
\begin{center}
\includegraphics[width=3.3in]{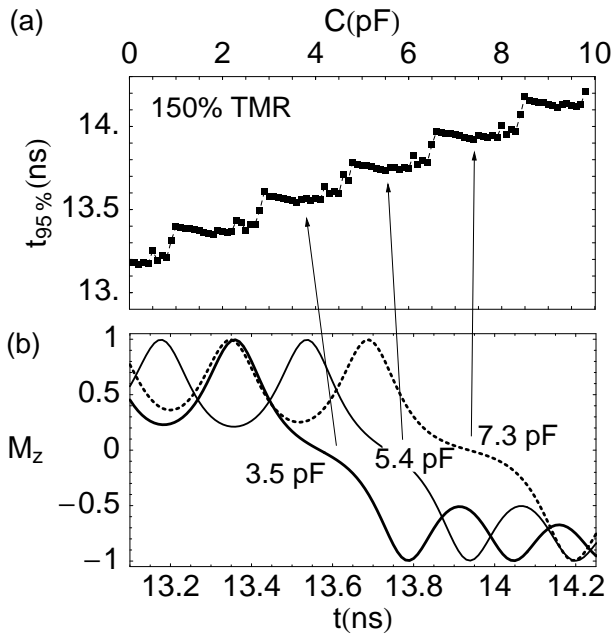}
\caption{(a) Switching time of a MTJ with $R_P$=2.5 k$\Omega$ and TMR=150$\%$ as a function of capacitance. (b) Easy axis component of the free layer magnetization as a function of time for succesive plateau minima located at $C_L$=3.5 pF (thick line), $C_L$=5.4 pF (narrow line), and $C_L$=7.3 pF (dashed line), detailing the switching process.}\label{fig:cap_effect}
\end{center}
\end{figure}

Now we consider the case $C_L\neq0$. Even though the capacitance of typical nano-scale MTJ's is of the order of fF, capacitance of the device leads can be orders of magnitude larger. Since the typical period of magnetization precession and current oscillations is in the GHz range, capacitive shunting becomes significant for k$\Omega$ device resistances. The effect of capacitance on the pulse duration required for 95$\%$ switching probability, $t_{95\%}$, for a MTJ with TMR=150$\%$ and R$_P$=2.5 k$\Omega$ is shown in Fig.~\ref{fig:cap_effect}(a). Here we do not consider any resistance feedback effects since the device resistance is large. The switching time is almost independent of capacitance below 0.5 pF, but increases non-monotonically and shows steps with similar height and width between 0.5 pF and 10 pF, the maximum studied capacitance value. However, the effects of capacitance are small: capacitance values close to 10 pF are required to increase $t_{95\%}$ from 13 ns to 14 ns. To explain the origin of the steps, magnetization trajectories at consecutive plateaus are shown in Fig.~\ref{fig:cap_effect}(b) for $C_L$=3.5 pF (wide line), $C_L$=5.4 pF (narrow line), and $C_L$=7.3 pF (dashed line). These traces show that as the capacitance is increased from one plateau to the next, reduction in the current through the device due to increased shunting slows the increase in precession amplitude to the point where an additional half-cycle is required for magnetization switching (e.g. for the dashed trace, switching occurs half a precession cycle after the switching point of the narrow trace, and one full precession cycle after the switching point of the thick line). However, once an additional half-cycle is required for switching, the slightly slower build-up of the magnetization precession amplitude caused by a slight increase in capacitance decreases the switching time, since the magnetization arrives to the ``switching'' half-cycle slightly earlier. This is observed in each of the steps of Fig.~\ref{fig:cap_effect}(a), where the switching time is larger at the beginning of the step and then decreases slightly towards the middle of the step. The observed decrease is only possible since the precession time increases with precession amplitude.

\section{Conclusions}

We studied the effect of resistance feedback on both spin valve and MTJ spin torque switching. Even though the critical current for instability is unchanged by this feedback (since the $\theta^2$ dependence of the resistance vanishes after the linearization procedure used for stability analysis~\cite{sun_PRB2000}), we determined how the presence of RIF and RVF affects the dynamical switching of spin valves and MTJ's. We observed that for spin valves in the $g(\theta,p)$=const. approximation the switching process in the presence of resistance-current feedback is asymmetric, with P-AP switching requiring lower voltage amplitudes and shorter duration pulse widths. We also observe that the amplitude of the oscillating spin torque saturates only for P-AP switching, explaining why in this case RIF produces the largest effects. Similar conclusions were obtained when using Slonczewski's ballistic $g(\theta,p)$ for spin valve switching, where the large difference in spin torque efficiency between $\theta\sim 0$ and $\theta\sim\pi$ increases the P-AP/AP-P switching asymmetry. For MTJ's with resistance comparable to the transmission line impedance we observe resistance-voltage feedback since, in contrast with spin valves, the torque is not current but voltage dependent. P-AP switching times are lowered by the presence of RVF whereas AP-P switching times are only slightly increased. As the MTJ resistance increases, the effect or RVF is reduced since the device voltage is almost unchanged by the time dependent magnetoresistance. Asymmetry in P-AP vs. AP-P switching times as large as 20$\%$ is expected to occur in large TMR, low RA MTJ's due to the existence of RVF.

We also studied the effect of capacitance on large resistance MTJ switching. We observed a stepped increase in the switching times as the effective device capacitance is increased. The overall change in switching times is small, e.g. only for capacitances above 10 pF does a $\sim$10$\%$ increase in $t_{95\%}$ occur. These results show that additional capacitance, such as that due to the circuitry used to measure and manipulate the state of the MTJ, will not significantly affect the MTJ switching time, as long as it is kept within the pF range. 

\begin{acknowledgements}

We thank D.C. Ralph for helpful comments and suggestions.

\end{acknowledgements}

\bibliography{references}

\end{document}